\documentclass[12pt]{iopart}

\usepackage{lineno,hyperref}
\modulolinenumbers[5]
\usepackage{iopams}
\usepackage{graphicx}
\usepackage{bm}% bold math
\usepackage{color}
\usepackage{xcolor}
\begin{document}

\title{New supercurrent pattern in quantum point contact with strained graphene nanoribbon}

\author[1]{Leyla Majidi}
\address{School of Nano Science, Institute for Research in Fundamental Sciences (IPM), P. O. Box 19395-5531, Tehran, Iran}
\ead{Leyla.majidi@ipm.ir}
\author[2]{Reza Asgari}
\address{School of Nano Science, Institute for Research in Fundamental Sciences (IPM), P. O. Box 19395-5531, Tehran, Iran}
\address{School of Physics, Institute for Research in Fundamental Sciences (IPM), P. O. Box 19395-5531, Tehran, Iran}
\address{ARC Centre of Excellence in Future Low-Energy Electronics Technologies, UNSW Node, Sydney 2052, Australia}

\vspace{10pt}
\begin{indented}
\item[]September 2020
\end{indented}

\begin{abstract}
We theoretically reveal the unusual features of the Josephson effect in a strained zigzag graphene nanoribbon with a small length relative to the superconducting coherence length and an arbitrary width. We find a step-wise variation of the critical supercurrent with the width of the nanoribbon, showing additional small width plateaus placed between the broad steps of a unstrained structure. We further demonstrate the peculiar quantization of the critical supercurrent in terms of the strain, resulted from the coupling of the pseudospin of Dirac fermions with the strain-induced gauge potential, where the height of the steps decreases with growing the strength of the fictitious gauge potential. Moreover, our results determine the potential of the proposed superconducting quantum point contact for the realization of the supercurrent switch under an applied strain. Besides, we find the local density of states of the strained zigzag nanoribbon displays a crossover between the decaying and oscillating behavior with the distance from the edges, by tuning the width and Fermi wavelength of the nanoribbon.
\end{abstract}

\maketitle

\section{Introduction}\label{sec:introduction}
The superconducting nanodevices are in the focus of modern experimental research, especially since they are a promising platform for various qubit realizations like Josephson-based qubits~\cite{Nakamura97,Han01,Vion02,Martinis02,Berkley03} or Majorana bound states~\cite{Kitaev01,Lutchyn10,Oreg10,Alicea12,Mourik12}. These structures containing superconductor-semiconductor and superconductor-insulator junctions host Andreev bound states which can also be used as a qubit~\cite{Janvier15,Chtchelkatchev03}. Andreev reflection~\cite{Andreev64} (successive conversion of electron-hole excitations at a normal-superconductor interface) is a subject of intensive theoretical and experimental research that spans far beyond quantum information topics~\cite{Blonder82,Deacon10,Lee13,Bretheau14,Tosi19}.

Fabrication of the nanoelectronic devices has provided the possibility of detecting the effects of electronic transport through a few or even single quantum states. The generic effect is the quantization of the conductance of a quantum point contact~\cite{van Wees88,Wharam88}. The analogous behavior was predicted~\cite{Beenakker91,Furusaki91} and experimentally confirmed~\cite{Takayanagi95} to occur for the Josephson current through an ordinary superconducting quantum point contact shorter than the superconducting coherence length.

Owing to the interplay of superconductivity and the unique electronic structure of graphene, graphene-based superconductors have attracted considerable attention in quantum transport and application of superconductor nanoelectronics. Many theoretical~\cite{Titov06,Gonzalez07,Hagymasi10,Black-Schaffer08,Linder08} and experimental works~\cite{Heersche07,Du08,Coskun12,Borzenets11} have been focused on superconductor-normal-superconductor nanostructure (Josephson junction) and other graphene-based superconducting heterostructures~\cite{Beenakker061,Cayssol08,Majidi12,Majidi13} and found peculiar and unexpected behaviors. Most of these features are the result of the massless Dirac spectrum of the low-lying electron-hole excitations in graphene, which in addition to the regular spin appear to come endowed with the two quantum degrees of freedom, the so-called pseudospin and valley~\cite{Novoselov04,Novoselov05,Zhang05}. These other degrees of freedom have been proposed separately to be used for controlling the electronic devices in pseudospintronics~\cite{Jose09,Majidi11,Majidi13} and valleytronics~\cite{Rycerz07}. To be specific, it has been demonstrated that a finite Josephson current can flow even in the limit of zero concentrations of the carriers at the Dirac point with a non-sinusoidal current-phase relation~\cite{Titov06}. Besides, a new class of graphene-based Josephson devices with novel properties has been realized for nanoribbons with various type of edges. In contrast to ordinary superconducting quantum point contact, the supercurrent in smooth or armchair edge with a low concentration of the carriers is unquantized and shows a monotonic decrease with lowering the width of the nanoribbon~\cite{Moghaddam06}. However, the zigzag type of the edges supports a half-integer quantization of the supercurrent, owing to the valley filtering nature of the zigzag graphene nanoribbon (zGNR) for the waves with mixed valley components~\cite{Moghaddam06}.

The prospect of using strain to engineer the electronic properties of graphene has opened up unprecedented opportunities and directions for graphene research \cite{vozmediano2016,vozmediano2010,asgari}. It has been shown that reversible and controlled strains in graphene can be realized by using a suitable substrate patterning~\cite{Ni08,Pereira09_1}, a uniform planer tension~\cite{Pereira09} or atomic force microscope tip~\cite{Lee08}. In the case of graphene being uniaxially strained, gapless graphene may turn to gapped graphene at critical strain~\cite{Choi10}. Owing to the large elastic deformation of graphene, the strain has the advantage of high tunability. The strained graphene can induce a valley-dependent pseudo-vector potential perpendicular to the direction of strain, due to shifted valley-dependent Dirac point in the strained region~\cite{Pereira09_1,Low10}. This leads to valley polarization, a significant characteristic of valleytronics. Number of such valley filters have been proposed in graphene and other two-dimensional materials~\cite{Fujita10,Rycerz07,Majidi14_2,Majidi14_1,Majidi14_0}. A giant pseudo-magnetic field greater than 300 Tesla resulting from the strongly deformed graphene was observed in graphene nanobubbles~\cite{Levy10}. Strain-induced pseudo-Landau levels have also been observed in graphene by chemical vapor deposition~\cite{Yeh11}.

Putting all these together, we came up with the idea of a novel class of Josephson junctions with the capability to sustain tunable charge transport in a zigzag-edged nanoribbon by means of mechanically induced strain.

Based on this idea, in this paper, we theoretically study the quantum transport in Josephson junction of a strained graphene nanoribbon with zigzag edges between two heavily doped superconductors. The coupling of the pseudospin of Dirac fermions with the strain-induced gauge potential produces a pseudo-magnetic field which has important effects on transport properties of the strained graphene. Making use of Dirac-Bogoliubov-de Gennes formalism~\cite{Beenakker061}, we find a peculiar quantization of the supercurrent in this kind of the Josephson junction with graphene nanoribbon of length $L$ smaller than the superconducting coherence length and an arbitrary width $W$. The special quantization relation of transverse momenta and the valley-filtering property of electron wave function in a zGNR make the supercurrent to be discretized in units of the even multiplies of the supercurrent quanta ($e\Delta_0/\hbar$) in an ordinary superconducting quantum point contact. The critical supercurrent variation with the width of the zGNR is different from that of a unstrained structure due to the presence of the strain-dependent small width plateaus in the middle of the broad unstrained plateaus. The width of the steps depends strongly on the strength of the applied strain ($\delta t/t_0$, with $t_0$ the nearest-neighbor hopping parameter in pristine graphene) and the Fermi wavelength inside the unstrained nanoribbon ($\lambda_F^0/a_0$, with $a_0$ the carbon-carbon distance for pristine graphene), such that the plateaus are getting broader and shifted to larger values of $\mu_0 W/h v_{\rm F}^0$ ($\mu_0$ and $v_{\rm F}^0$ are respectively the chemical potential and Fermi velocity inside the pristine graphene nanoribbon) with increasing the $\delta t/t_0$ and $\lambda_F^0/a_0$ values. Also, the height of the steps is getting reduced by growing at the temperature of the device. Moreover, we demonstrate that the Josephson current can be switched on/off for small values of $\mu_0 W/ h v_{\rm F}^0$ because of the displacement of the Dirac points in the presence of applied strain.

We further present the peculiar behavior of the critical supercurrent in terms of the fictitious gauge potential (or the pseudo-magnetic field). Depending on the value of the $\lambda_F^0/a_0$ ratio, the critical supercurrent possesses a constant value or asymmetric step-wise behavior with a wide peak around zero pseudo-magnetic fields. The width of the peak, and also the height and width of the plateaus reduce by enhancing the value of $\lambda_F^0/a_0$ and $\mu_0 W/h v_{\rm F}^0$ ratios. Depending on the Fermi wavelength inside the pristine normal graphene nanoribbon, the Josephson current can be turned off for an accessible experimental range of the applied strain. In addition, the current-phase relation is found to be similar to that of an ordinary superconducting quantum point contact.

In addition, we investigate the behavior of the local density of states (LDOS) inside the strained zGNR. The position dependence of the LDOS provides us more information about the edge states supported by the zGNR. Depending on the value of $\lambda_F^0/a_0$ and $\mu_0 W/h v_{\rm F}^0$ ratios, the LDOS at zero excitation energy may have equal high values at the opposite edges resulting from the evanescent mode confined to the interface and decreases with the distance from the edges or it may have oscillatory behavior with different values at the edges resulting from the propagating states. The finite excitation energy leads to an increase in the local density of states at the edge $y=0$ and in decrease at the opposite edge $y=W$. The surface states at the opposite edges of the ribbon with the positions $y=0$ and $y=W$ are those coming from the A and B sublattices, respectively.

The paper is organized as follows. In Sec. \ref{sec:model}, we introduce the model and establish the theoretical framework which will be used to investigate the Josephson effect and the LDOS in a strained graphene nanoribbon with zigzag edges. Then, in Sec. \ref{sec:supercurrent} the critical supercurrent dependence on the width of the nanoribbon and also on the fictitious gauge potential are shown, followed by a discussion over the importance of the results. Section \ref{sec:LDOS} is devoted to the results of the local density of states inside the strained normal nanoribbon with zigzag edges. Finally, our conclusions are summarized in Sec. \ref{sec:conclusion}.

\section{Model and basic equations}\label{sec:model}
\begin{figure}[]
\begin{center}
\includegraphics[width=4.5in]{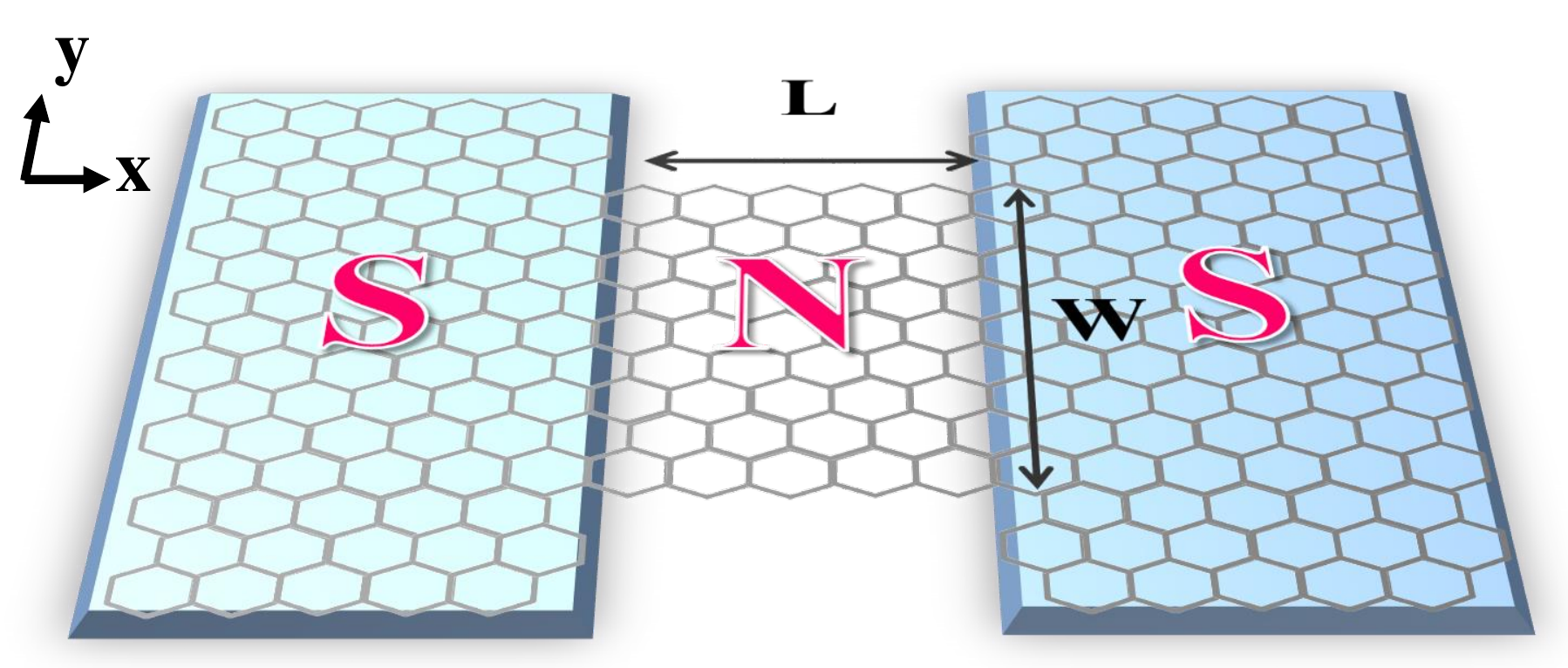}
\end{center}
\caption{\label{Fig:1} Schematic illustration of the graphene-based superconducting quantum point contact; the two superconducting S regions (caused by proximity to S electrodes) are connected through an in-plane strained zGNR of length $L$ and width $W$.}
\end{figure}
The device we consider is a Josephson junction between two s-wave superconducting (S) electrodes deposited on top of a graphene sheet, which are separated by a zigzag normal (N) graphene nanoribbon of length $L$ [see Fig. \ref{Fig:1}]. The N region is imposed on a uniform planar tension which can lead to uniform fictitious gauge potential.

The effects of strain on the electronic properties of graphene are usually captured by utilizing a nearest-neighbor tight-binding model. Taking into account all strain-induced effects: changes in the nearest-neighbor hopping parameters, the reciprocal lattice deformation and the true shift of the Dirac cones, the effective Dirac Hamiltonian can be written as~\cite{Oliva,Vozmediano}
\begin{equation}
\label{H}
\mathcal{H}^{\tau}=v_{\rm F}\bm{\sigma}^{\tau}\cdot\bm{p}-\tau \hbar v_{\rm F}^0\ \bm{\sigma}^{\tau}\cdot\bm{A}_s
\end{equation}
for the K(K') valley with $\tau=1 (-1)$, where $\bm{\sigma}^{\tau}=(\sigma_x,\tau \sigma_y)$ is the vector of the Pauli matrices operating in the sublattice or pseudospin space, and $\bm{p}=(p_x,p_y)$ is the two-dimensional momentum. Here, $v_{\rm F}^0$ represents the Fermi velocity in pristine graphene, and its strained counterpart $v_{\rm F}$ ($v_{\rm F}=\sqrt{v_x v_y}$) can be obtained using
\begin{equation}
\bm{v}=(\hat{I}+\hat{\epsilon}-\beta\hat{\epsilon})\cdot\bm{v}^0,
\end{equation}
where $\hat{I}$ is the $2\times2$ identity matrix, $\hat{\epsilon}$ is the position-independent strain tensor and $\beta=|{\partial\ln{t}}/{\partial\ln{a_0}}|\simeq 2$ is the electron Gr$\ddot{u}$neisen parameter. The $\beta$-dependence term is owing to the strain-induced changes in the hopping parameter and its contribution depends on the material since $\beta$ varies depending on the material. Here and below, the index zero denotes the absence of strain. The strain-induced gauge potential $\bm{A}_s$ is also related to the strain tensor by
\begin{eqnarray}
A_{s,x}&=&\frac{\beta}{2 a_0}(\epsilon_{xx}-\epsilon_{yy}),\nonumber\\
A_{s,y}&=&-\frac{\beta}{2 a_0}(2\epsilon_{xy}).
\end{eqnarray}
 The two elements $\epsilon_{xx}$ and $\epsilon_{yy}$ of the strain tensor are the axial strains and $\epsilon_{xy}$ is the shear strain. Applying a uniform tension along the direction with an angle $\theta$ with respect to $x$-axis (zigzag direction) results in a strain tensor~\cite{Pereira09}
\begin{equation}
\hat{\epsilon}=\epsilon\ \left(
\begin{array}{cc}
{\cos}^2{\theta}-\nu\ {\sin}^2{\theta} & (1+\nu)\sin{\theta}\cos{\theta} \\
(1+\nu)\sin{\theta}\cos{\theta}& {\sin}^2{\theta}-\nu\ {\cos}^2{\theta}\\
\end{array}
\right),
\end{equation}
where $\epsilon$ is the tensile strain and $\nu=0.165$ is the Poisson's ratio.

The strain-induced gauge potential $\bm{A}_s$ and the corresponding field $\bm{B}_s$ ($\bm{B}_s=\nabla\times \bm{A}_s$) have opposite signs at the vicinity of the two Dirac points (K and K') at the corners of the hexagonal first Brillouin zone. The underlying deep reason for the sign change originates from the fact that strain and any geometric deformation do not break time-reversal symmetry, and the two $K$ points or valleys are connected by time-reversal operator ($K\leftrightarrow K'$). Explicitly, we would like to have $\bm{A}_s=(\hbar v_{\rm F}^0)^{-1}\delta t [\Theta(x+L/2)-\Theta(x-L/2)]\hat{\bm y}$, where $\delta t$ parameterizes the strain by its effect on the nearest-neighbor hopping $t_0\rightarrow t_0+\delta t$ ($t_0\simeq 3$ eV). This corresponds to the tensional strain along $\theta=-\pi/4$ direction with
$\delta t= 3 \beta t_0 \epsilon (1+\nu)/4$, and produces the isotropic transport with modified Fermi velocity
\begin{equation}
\label{velocity}
v_{\rm F}=v_{\rm F}^0\ [1+(1-\beta)(\epsilon_{xx}+\epsilon_{xy})].
\end{equation}

Therefore, the cross section of the energy dispersion is circular, and the chemical potential of the strained N region is equal to $\mu=\hbar v_{\rm F} k_{\rm F}$, with $k_{\rm F}=\sqrt{n\pi}$ and $n$ the electron density. The strain-induced modification of the lattice primitive vectors in the form of $\bm{a}_i=(\hat{I}+\hat{\epsilon})\cdot\bm{a}_i^0$~\cite{Pereira09}, with $\bm{a}_{1(2)}^0=(\pm1,\sqrt{3})\sqrt{3}a_0/2$, leads to the changes in the primitive cell area as $\Omega=\Omega_0(1+\epsilon_{xx}+\epsilon_{yy}+\epsilon_{xx} \epsilon_{yy}-\epsilon_{xy}^2)$, and accordingly in the electron density
as~\cite{{Rostami}}
\begin{equation}
\label{electron denisty}
n=\frac{n_0}{1+\epsilon_{xx}+\epsilon_{yy}+\epsilon_{xx} \epsilon_{yy}-\epsilon_{xy}^2}.
\end{equation}
Using the above-found relations [Eqs. (\ref{velocity}) and (\ref{electron denisty})], the modified chemical potential up to first order in strain tensor can be written as
\begin{equation}
\mu=\mu_0\ \frac{[1+(1-\beta)(\epsilon_{xx}+\epsilon_{xy})]}{\sqrt{1+\epsilon_{xx}+\epsilon_{yy}}},
\end{equation}
and accordingly the modified Fermi wavelength will be obtained as
\begin{equation}
\lambda_F=\lambda_F^0 \sqrt{1+\epsilon_{xx}+\epsilon_{xy}},
\end{equation}
with $\mu_0=\hbar v_{\rm F}^0 \sqrt{n_0\pi}$ and $\lambda_F^0=h v_F^0/|\mu_0|$ the chemical potential and Fermi wavelength inside the pristine graphene, respectively.

Inside the part of graphene covered by the S electrode, an effective potential $\Delta$ is induced via proximity effect. The S parts are assumed to be heavily doped such that the Fermi wavelength inside them is much smaller than the superconducting coherence length $\xi$ ($\xi=\hbar v_{\rm F}^0/|\Delta|$) and also the Fermi wavelength in the N graphene nanoribbon to justify the mean-field theory of superconductivity and neglect the spatial variation of the pair potential $\Delta$ inside the superconductors close to the N/S interfaces.

The superconducting correlation between relativistic electrons and holes with opposite spins and different valley indices can be described by Dirac-Bogoliubov-de Gennes (DBdG) equation~\cite{Beenakker061}
\begin{equation}
\label{DBdG}
\left(
\begin{array}{cc}
\mathcal{H}^{\tau}-\mu-U(\bm{r}) & \Delta \\
\Delta^{\ast}& \mu+U(\bm{r})-\mathcal{H}^{\tau}
\\
\end{array}
\right)
\left(
\begin{array}{c}
u_{\tau}\\
v_{\bar{\tau}}
\end{array}
\right)
=\varepsilon\left(
\begin{array}{c}
u_{\tau}\\
v_{\bar{\tau}}
\end{array}
\right).
\end{equation}
Notice that unstrained parameters are used for S regions and the strain-induced gauge potential $\bm{A}_s$ is zero in S regions. Here, $\varepsilon$ is the excitation energy, $U(\bm{r})$ is the electrostatic potential taken to be $U_0\gg \mu_0$ in S regions and zero in N region, and $\bar{\tau}=-\tau$. Also, the electron and hole wave functions, $u_{\tau}$ and $v_{\bar{\tau}}$, are two-component spinors of the form $(\psi_A,\psi_B)$, where the two components give the amplitude of the wave function on the two sublattices. Therefore, the electron excitations in one valley are coupled by the superconducting pair potential $\Delta$ to hole excitations in the other valley.

We assume a phase difference $\phi$ between two S regions, with pairing functions $\Delta_{L,R}=\Delta_0 e^{{\pm i\phi/2}}\sigma_0$, to drive a Josephson supercurrent through the graphene nanoribbon which constitutes a week link between two superconductors. The Josephson current can be obtained from the formula~\cite{Beenakker92}
%\begin{widetext}
\begin{eqnarray}
\label{supercurrent}
\hspace{-1cm}I(\phi)&=&-\frac{2 e}{\hbar}\sum_p \tanh(\frac{\varepsilon_p}{2 k_B T}) \frac{d\varepsilon_p}{d\phi} -\frac{2 e}{\hbar} (2k_B T)\int_{\Delta_0}^{\infty} d\varepsilon \ln[2\cosh(\frac{\varepsilon}{2k_B T})] \frac{\partial\rho}{\partial\phi}\nonumber\\
\hspace{-1cm}&+&\frac{2 e}{\hbar} \frac{d}{d\phi}\int d\bm{r} \frac{|\Delta|^2}{|g|},
\end{eqnarray}
%\end{widetext}
where the factor of $2$ accounts for the two-fold spin degeneracy, $\rho$ is the density of state and $g$ is the interaction coefficient of the BCS theory of the superconductivity. For the step function model of $\Delta(\bm{r})$, $|\Delta|$ is independent of $\phi$ so that the contribution of the third sentence can be disregarded. A calculation of the Josephson current subsequently requires a unique knowledge of the eigenvalues. For experimentally relevant short junction regime $L\ll\xi$, the bound (discrete) states with energies $\varepsilon_p<\Delta_0$ have the main contribution to the Josephson current. In this case, contributions from continuous states with energies $\varepsilon\geq\Delta_0$ may be neglected. Therefore, the Josephson supercurrent will be carried by the so-called subgap Andreev bound states, formed in the N region owing to the Andreev reflection at the N/S interfaces.

This local coupling of the electron and hole excitations at an ideal N/S interface (at a point $\bm{r}$) can be described by means of a longitudinal boundary condition on the electron and hole wave functions in the N region~\cite{Titov06}
\begin{equation}
\psi_h(\bm{r})=e^{-i\Phi-i\beta' \hat{n}\cdot\bm{\sigma}}\psi_e(\bm{r}),
\end{equation}
where $\Phi$ is the phase of the pair potential $\Delta$ in S region, $\hat{n}$ is a unit vector perpendicular to the interface pointing from N to S region, and $\beta'=\arccos{(\varepsilon_p/\Delta_0)}$.

In the proposed S/N/S structure with superconducting phases $\Phi_{L,R}=\pm \phi/2$, the boundary condition at the two interfaces $\bm{r}_{\mp}=(\mp L/2,y)$ takes the form~\cite{Titov06}
\begin{equation}
\psi_h(\bm{r}_{\mp})={U^{\pm 1}(\varepsilon_p,q)}\psi_e(\bm{r}_{\mp})
\end{equation}
with $U(\varepsilon_p,q)=e^{-i\phi/2}e^{i\beta' \sigma_x}$, while the electron (hole) states at the two ends can be related via the transfer matrix,
\begin{equation}
\psi_{e(h)}(\bm{r}_{+})=M(\pm \varepsilon_p,q)\psi_{e(h)}(\bm{r}_{-}).
\end{equation}
The transfer matrix $M(\pm \varepsilon_p,q)$ can be obtained from Eq. (\ref{H}) as
\begin{eqnarray}
&&M(\pm \varepsilon_p,q)={M_s^{-1}(\pm\varepsilon_p,q)}\ M_0(k,L)\ M_s(\pm \varepsilon_p,q),\\
&&M_s(\pm \varepsilon_p,q)=\frac{1}{\sqrt{2 \cos{\alpha}}}\left(
\begin{array}{cc}
e^{-i \alpha/2} & e^{i \alpha/2} \\
e^{i \alpha/2}& -e^{-i \alpha/2}\\
\end{array}
\right),\\
&&M_0(k,L)=e^{ikL\sigma_z},
\end{eqnarray}
where $\alpha=\arcsin{[(\hbar v_{\rm F} q-\delta t)/(\mu+\varepsilon_p)]}$ indicates the angle of the propagation of the quasiparticles inside the N region at a transverse wave vector $q$ with longitudinal wave vector $k=(\hbar v_{\rm F})^{-1}\sqrt{(\mu+\varepsilon_p)^2-(\hbar v_{\rm F} q-\delta t)^2}$.
Applying the condition of the constructive interference for the wave function, after a round trip from $\bm{r}_-$ to $\bm{r}_+$ and back to $\bm{r}_-$, leads to the usual form of the Andreev bound energies in terms of the normal-state transmission probability $T_P$ in the short junction limit $L\ll \xi$,
\begin{equation}
\varepsilon_p(\phi)=\Delta_0 \sqrt{1-T_p\ \sin^2(\frac{\phi}{2})}.
\end{equation}
Substituting the above-found relation into Eq. (\ref{supercurrent}) yields the Josephson current as
%\begin{widetext}
\begin{equation}
I(\phi)=\frac{e\Delta_0}{2\hbar}\sum_p \frac{T_p \sin{\phi}}{\sqrt{1-T_p\ {\sin^2(\frac{\phi}{2})}}}\ \tanh[\frac{\Delta_0}{2 k_B T}\sqrt{1-T_p\ \sin^2(\frac{\phi}{2})}].
\end{equation}
%\end{widetext}
To evaluate the transmission probability ($T_p$), we consider the proposed structure in the normal state (with $\Delta_0=0$); the strained zGNR between two heavily doped N regions. The total wave function of the right (left) going wave inside the strained zGNR is composed of the pseudospinor wave functions with transverse momentum $\pm q$ from two valleys
\begin{equation}
\Psi^{\pm}=A\ \psi_q^{k\pm}+B\ \psi_{-q}^{k\pm}+A'\ \psi_q^{'k'\pm}+B'\ \psi_{-q}^{'k'\pm},
\end{equation}
where $ \psi_{\pm q}^{k\pm} $ and $ \psi_{\pm q}^{'k'\pm}$ are respectively the solutions of Dirac equation [Eq. (\ref{H})] for the K and K' valleys and the two propagation directions along the $x-$axis are denoted by $\pm$ in $\Psi^{\pm}$.

Imposing the boundary condition of zigzag edges~\cite{Brey06}, $\Psi_A(y=0)=\Psi'_A(y=0)=\Psi_B(y=W)=\Psi'_B(y=W)=0$, leads to two sets of wave functions in the space of pseudospin and valley degrees of freedom; a right (left) going wave in the K(K') valley
\begin{eqnarray}
\Psi^{+}&=&\ e^{ikx}\left(
  \begin{array}{c}
 \sin{qy} \\
 e^{i(\alpha-\alpha')/2}\ \sin[{qy+(\frac{\alpha+\alpha'}{2})}]\\0\\0\\
  \end{array}
\right),\nonumber\\\\
\Psi^{'-}&=&\ e^{-ik'x}\left(
  \begin{array}{c}
0\\0\\ \sin{qy} \\
- e^{-i(\alpha-\alpha')/2}\ \sin[{qy+(\frac{\alpha+\alpha'}{2})}] \\
  \end{array}
\right),\nonumber\\
\end{eqnarray}
 with transcendental relation $\sin[{qW+(\alpha+\alpha')/2}]=0$ for allowed values of $q$, and a left (right) going wave in the $K(K')$ valley
 \begin{eqnarray}
\Psi^{-}&=&\ e^{-ikx}\left(
  \begin{array}{c}
\sin{qy} \\
-e^{-i(\alpha-\alpha')/2}\ \sin[{qy-(\frac{\alpha+\alpha'}{2})}] \\0\\0\\
  \end{array}
\right), \nonumber\\
\Psi^{'+}&=&\ e^{ik'x}\left(
  \begin{array}{c}
0\\0\\ \sin{qy} \\
e^{i(\alpha-\alpha')/2}\ \sin[{qy-(\frac{\alpha+\alpha'}{2})}] \\
  \end{array}
\right), \nonumber\\
\end{eqnarray}
with transcendental relation $\sin[{qW-(\alpha+\alpha')/2}]=0$, where $\alpha'=\arcsin{[(\hbar v_{\rm F} q+\delta t)/(\mu+\varepsilon_p)]}$ and $k'=(\hbar v_{\rm F})^{-1}\sqrt{(\mu+\varepsilon_p)^2-(\hbar v_{\rm F}q+\delta t)^2}$. Note that these transcendental relations couple the transverse momentum $q$ to the excitation energy $\varepsilon_p$, and have finite number of solutions (allowed quantized $q$) depending on the value of $\mu_0 W/h v_{\rm F}^0$ [see Fig. \ref{Fig:3}(a)].

Most importantly, the above-found wave functions for the right and left going waves inside the N region notifies us that for each allowed mode, zGNR operates as a valley-filter for the waves with mixed valley components. This filtering property prevents the electrons from normal reflection, due to the conservation of the transverse momentum $q$ and the valley index $\tau$. Consequently, perfect transmission (with $T_p=1$) occurs for right going waves with allowed quantized transverse momentum $q$.

Therefore, the valley filtering nature of the zGNR makes the Andreev bound states energy not to depend on the transverse mode $q$ and contains $N$-fold degenerate states at energy $\varepsilon_p=\Delta_0 |\cos{(\phi/2)}|$. The integer $N$ is the number of allowed transverse modes at the Fermi level propagating through the construction, which can be obtained from transcendental relations. Therewith, the current-phase relation becomes

\begin{equation}
\label{Iphi}
I(\phi)=N\frac{e\Delta_0}{\hbar}\sin(\frac{\phi}{2})\ \frac{\cos(\frac{\phi}{2})}{|\cos(\frac{\phi}{2})|}\ \tanh[\frac{\Delta_0}{2 k_B T}\cos(\frac{\phi}{2})],
\end{equation}
which is similar to the characteristic relation of an ordinary superconducting quantum point contact with the critical supercurrent $I_C=N e\Delta_0/\hbar$. Therefore, the calculation of the critical supercurrent requires the identification of the number of active transverse modes $N$ at a given energy (Fermi level), which can be obtained from the solutions of transcendental relations $\sin[{qW\pm(\alpha+\alpha')/2}]=0$.

The unique band structure of graphene may lead to the significant transmission of the evanescent modes in graphene nanoribbons.  In Josephson junctions with smooth- or armchair-edge graphene nanoribbon, the significant contribution of the evanescent modes relative to the propagating modes leads to the absence of supercurrent quantization~\cite{Moghaddam06}. In Josephson junction with zigzag-edge nanoribbon, on the other hand, the presence of atoms from different sublattices at the two edges results in a special quantization relation of transverse momenta with a finite number of solutions, depending on the value of the $\mu_0 W/h v_F^0$ ratio. All possible modes are propagating and hence, there is no evanescent mode. Therefore, the transport of the finite number of propagating modes leads to the quantization of the critical supercurrent in Josephson junction with zGNR.

Another experimentally accessible quantity is the local density of states (LDOS), which can be obtained through the following formula \cite{Gennes89},
\begin{equation}
\label{LDOS}
\tilde{N}(E,r)=\sum_{\bm{k}}{|\psi_{\bm{k}}(r)|^2\ \delta(E(\bm{k})-E)},\\
\end{equation}
where $\psi_{\bm{k}}(r)$ corresponds to the eigenfunction of energy $E(\bm{k})$ [$E(\bm{k})=\mu+\varepsilon(\bm{k})$] and the sum is over all states with the wave vectors $\bm{k}$.

To evaluate the LDOS inside the strained N region, we replace $\psi_{\bm{k}}(r)$ with the wave functions of the Andreev bound states for the incoming electrons from the K and K' valleys,
\begin{equation}
\psi_N=\left(
  \begin{array}{c}
 \Psi^{+}(\varepsilon_p)+r_e\ \Psi^{-}(\varepsilon_p)\\r_h\Psi^{'-}(-\varepsilon_p)
  \end{array}
\right),
\end{equation}
and
\begin{equation}
\psi_N^{'}=\left(
  \begin{array}{c}
 \Psi^{'+}(\varepsilon_p)+r'_e\ \Psi^{'-}(\varepsilon_p)\\r'_h \Psi^{-}(-\varepsilon_p)
  \end{array}
\right),
\end{equation}
respectively. We emphasize that the normal reflection of the incoming electron from the K(K') valley with the amplitude $r_e$ ($r'_e$) will be forbidden due to the conservation of the transverse momentum $q$ and the valley degree of freedom, and the electron-hole conversion (Andreev reflection process) with unit amplitude $r_h=1$ ($r'_h=1$) will occur. The numerical results discuss in the following section.

\section{Numerical results and discussion}\label{sec:results}

In this section, we present our numerical results for the critical supercurrent and the LDOS inside the uniform strained normal zGNR, using Eqs. (\ref{Iphi}) and (\ref{LDOS}). We scale the strain $\delta t$ in terms of the unstrained nearest-neighbor hopping parameter $t_0$, and the width of the zigzag nanoribbon $W$ in terms of the Fermi wavelength in pristine graphene $\lambda_F^0=hv^{0}_{\rm F}/|\mu_0|$ which itself can be expressed in terms of the equilibrium lattice constant $a_0$. In addition, we scale the temperature $T$ in terms of the critical temperature of the superconducting order parameter $T_C$. We set the superconducting order parameter $\Delta_0=1$\ meV and scale the excitation energy $\varepsilon_p$ in units of $\Delta_0$. We note the results are in the experimentally most relevant short junction regime ($L\ll\xi$). In terms of energy scales, this condition requires $\Delta_0\ll\hbar v_{\rm F}^0/L$. We emphasize that the applied strain ratio $\delta t/t_0$ is up to $0.3$ to persist the semimetallic nature of graphene~\cite{Choi10}.

\subsection{Critical supercurrent}\label{sec:supercurrent}

First, we evaluate the critical supercurrent in the proposed Josephson junction at zero temperature ($T=0$) by making use of Eq. (\ref{Iphi})
\begin{equation}
\label{Iphi0}
I(\phi)=I_C\sin(\frac{\phi}{2})\ \frac{\cos(\frac{\phi}{2})}{|\cos(\frac{\phi}{2})|},
\end{equation}
where the critical supercurrent is given by $I_C= max[I(\phi)]=Ne\Delta_0/\hbar$.

\begin{figure}[]
\begin{center}
\includegraphics[width=4.5in]{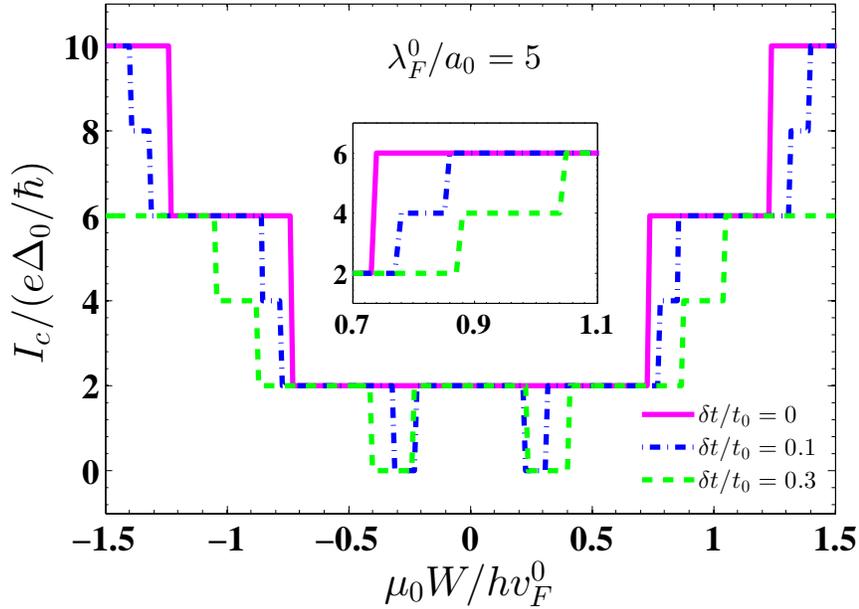}
\end{center}
\caption{\label{Fig:2} The critical supercurrent through the graphene-based Josephson junction with uniform in-plane strained zigzag-edge N nanoribbon versus the width of the N region $\mu_0 W / hv_{\rm F}^0$ for different values of $\delta t/t_0$, when $\lambda_F^0 / a_0=5$.
Note that the critical supercurrent increases step-wise as a function of the constriction width. In contrast to the unstrained case, the quantization is no longer half-integer. The applied strain supports new plateaus with small width at the middle of the wide unstrained plateaus. Inset presents the zoomed-in view of the critical supercurrent in the range $0.7\leq\mu_0 W / hv_{\rm F}^0\leq 1.1$.}
\end{figure}

\begin{figure}[]
\begin{center}
\includegraphics[width=4.5in]{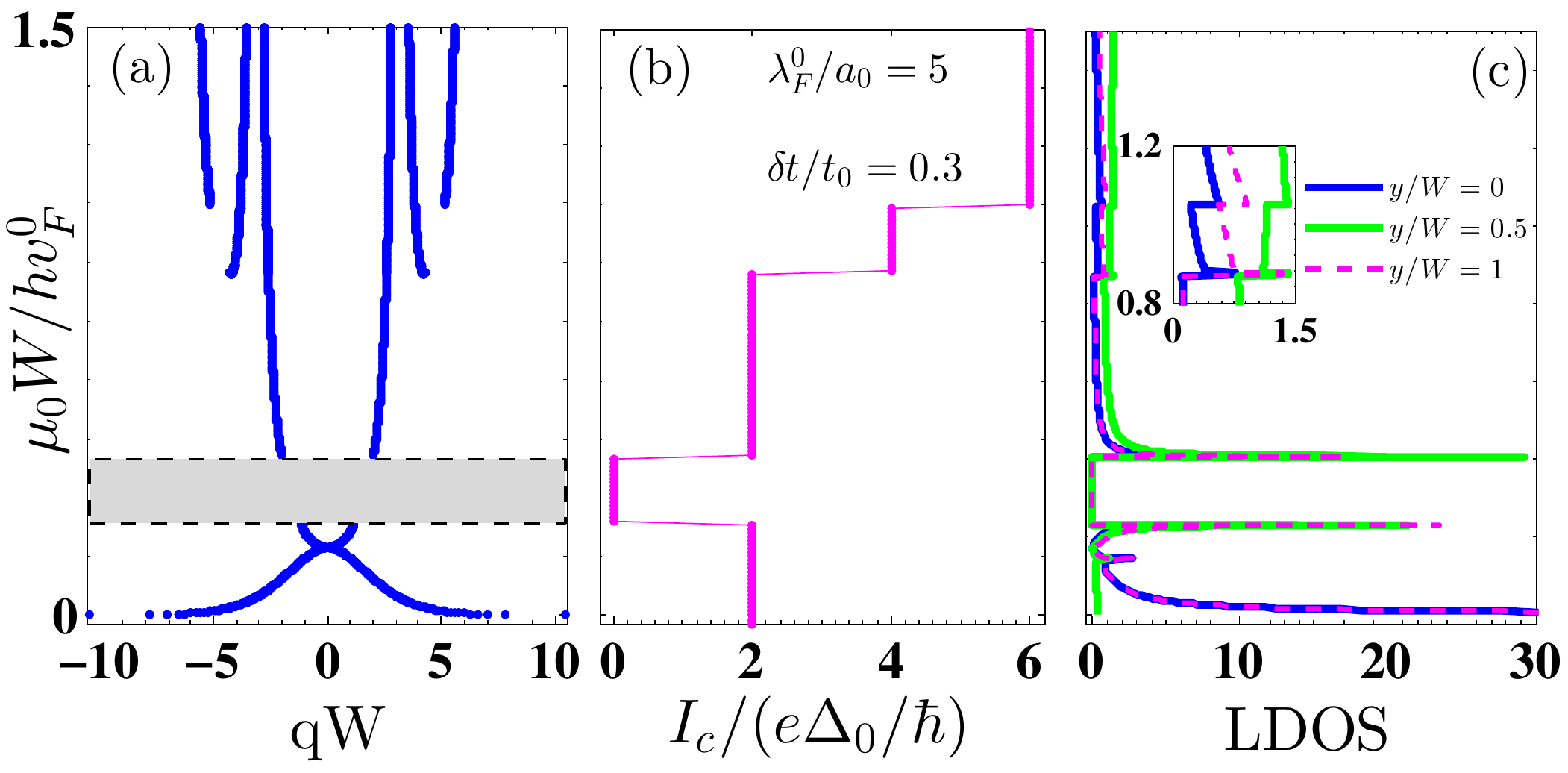}
\end{center}
\caption{\label{Fig:3} (a) The dimensionless width $\mu_0 W/h v_{\rm F}^0$ dependence of the quantized transverse wave vector $qW$ for $\delta t/t_0 = 0.3$, $\lambda_F^0/a_0 = 5$, and the excitation energy $\varepsilon_p/\Delta_0=0$. (b) and (c) correspond to the critical supercurrent and the local density of states (LDOS), respectively. The LDOS is presented for three different values of the dimensionless distance $y/W$ from the edges of the ribbon, when $x/L = 0.1$, and $L/W = 0.1$. Inset of (c) shows the zoomed-in view of the LDOS for the range $0.8\leq\mu_0 W/h v_{\rm F}^0\leq 1.2$. The gray region in (a) belongs to the non-transport regime. The step-wise behavior of the critical supercurrent is attributed to the number of the active transport channels in the proposed Josephson junction in which they carry the quantized supercurrent $e\Delta_0/\hbar$. Appearance of a flat band located at small $\mu_0 W/h v_{\rm F}^0$ value, gives rise to a strong LDOS peak at the edges of the strained zGNR with $y/W =0$ and $1$. Especially, the quantized level singularities (finger print of the supercurrent plateaus) are clearly observed accompanied with the vanishing LDOS for the $\mu_0 W/h v_{\rm F}^0$ values in the gray region.}
\end{figure}

Figure \ref{Fig:2} shows the behavior of the critical supercurrent $I_C$ in terms of the dimensionless width of the zGNR $\mu_0 W/h v_{\rm F}^0$ for different values of $\delta t/t_0$ when $\lambda_F^0 / a_0=5$. We mention that the critical supercurrent of the Josephson junctions with n- and p-doped N regions are combined in the same plot respectively with the positive and negative $\mu_0 W/h v_{\rm F}^0$ ratios, since $\mu_0 W/h v_{\rm F}^0=sgn(\mu_0) W/\lambda_F^0$. As seen here the critical supercurrent increases step-wise as a function of the constriction width, independent of the properties of the junction. For unstrained Josephson junction, this type of the edges supports a half-integer quantization of the supercurrent to $(n+1/2)4e\Delta_0/\hbar$, similar to the result of Ref.~\cite{Moghaddam06} and in contrast to an ordinary superconducting quantum point contact. This quantization of the critical supercurrent is unique to the zigzag-edge nanoribbon and can be unobserved for smooth or armchair edge nanoribbons~\cite{Moghaddam06}.

Imposing a uniform planar strain to the zGNR leads to the following important anomalies for the critical supercurrent. In contrast to the unstrained case, the quantization is no longer half-integer. The applied strain supports new plateaus with small width at the middle of the wide unstrained plateaus, such that their width enhances by increasing the $\delta t/t_0$ ratio. Also, the main wide plateaus are shifted to larger values of $\mu_0 W/h v_{\rm F}^0$. Most importantly, a supercurrent switching is seen in the lowest plateau where the applied strain can turn on/off the Josephson current depending on the values of $\delta t/t_0$ and $\mu_0 W/h v_{\rm F}^0$. This is attributed to the strain-induced displacement of the Dirac points and propose the Josephson junction under uniform planar strain as a supercurrent switch. We mention that the finite value of the critical supercurrent around $W= 0$ is a numerical artifact and it should be zero. This artifact is related to the finite value of $\mu_0 W/h v_{\rm F}^0= W/\lambda_F^0=10^ {-4} $.

Figure \ref{Fig:3} manifests that the step-wise behavior of the critical supercurrent is attributed to the number of the active transport channels in the proposed Josephson junction in which they carry the quantized supercurrent $e\Delta_0/\hbar$. In the regime of our interest, $\varepsilon_p<\Delta_0\ll\mu_0$, the number of allowed modes depends on the value of the $\mu_0 W/h v_{\rm F}^0$ ratio. As represented in Fig. \ref{Fig:3}(a), the strained graphene nanoribbon with zigzag edges has two bands crossing at small values of $\mu_0 W/h v_{\rm F}^0$ ratio (and finite momentum). Increasing the $\mu_0 W/h v_{\rm F}^0$ ratio leads to the enhancement of the number of available modes by two. Therefore, plateaus of the quantized critical supercurrent are observed at even multiples of $e\Delta_0/\hbar$ [see Fig. \ref{Fig:3}(b)]. The first step has a bigger width than that of the higher steps, due to the contribution of the single $y$-dimensional evanescent mode [depicted by the lowest branch in Fig. \ref{Fig:3}(a)]. Specially, the absence of transport modes for the values of the $\mu_0 W/h v_{\rm F}^0$ ratio in the gray region [see Fig. \ref{Fig:3}(a)] leads to the absorbing observation of the supercurrent switching. The effect resembles the conductance quantization in strained zGNR, where the clean quantum Hall plateaus are observed with filling factors given by $\nu'=2, 6, 10, ...=4n+2$~\cite{Low10,Rostami13,Diniz}.

\begin{figure}[]
\begin{center}
\includegraphics[width=4.5in]{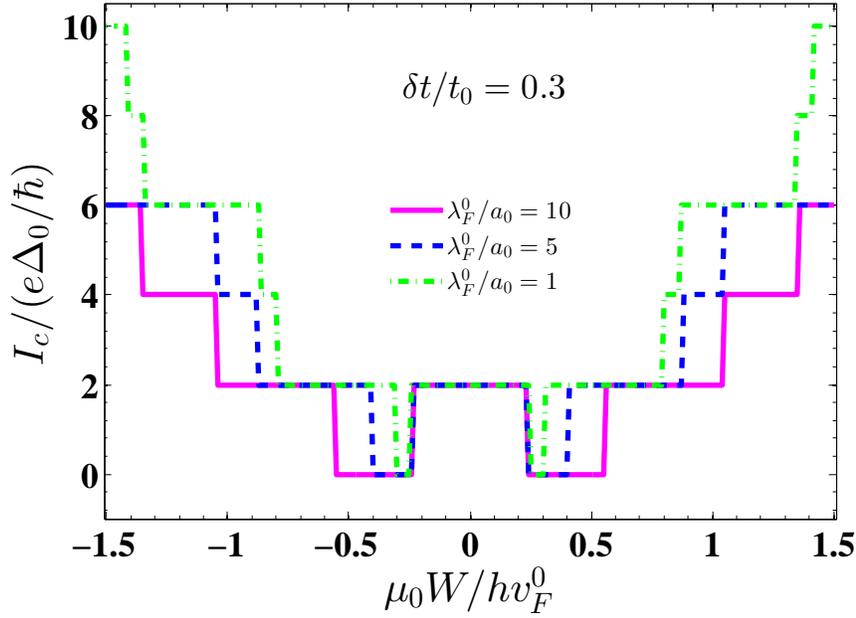}
\end{center}
\caption{\label{Fig:4} The behavior of the critical supercurrent $I_C$ in terms of the width of the N region $\mu_0 W / hv_{\rm F}^0$ for different values of $\lambda_F^0 / a_0$, when $\delta t/t_0=0.3$. Notice that decreasing the $\lambda_F^0 / a_0$ ratio, leads to a reduction of the sub-plateaus width with respect to $\mu_0 W/h v_{\rm F}^0$.}
\end{figure}

\begin{figure}[]
\begin{center}
\includegraphics[width=4.5in]{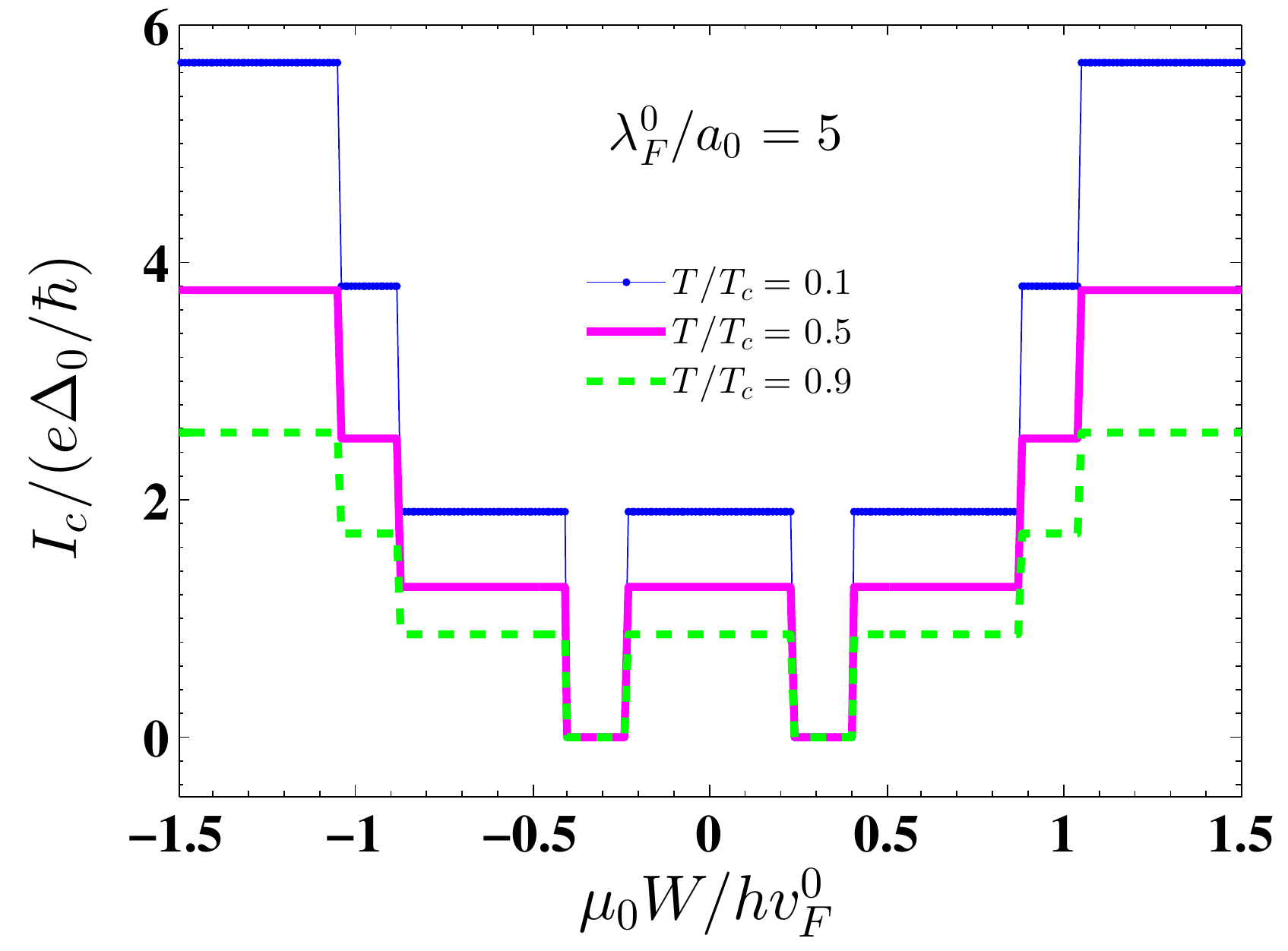}
\end{center}
\caption{\label{Fig:9} The critical supercurrent in terms of the width of the strained zigzag-edge nanoribbon $\mu_0 W/h v_{\rm F}^0$ for different values of the temperature $T/T_C$ ratio, when $\lambda_F^0 / a_0 = 5$ and $\delta t/t_0 = 0.3$. The step-wise variation of the critical supercurrent is preserved for finite-temperatures below $T_c$.}
\end{figure}

Moreover, it is demonstrated in Fig. \ref{Fig:4} that decreasing the $\lambda_F^0 / a_0$ ratio leads to a reduction of the sub-plateaus width with respect to $\mu_0 W/h v_{\rm F}^0$. The effect of finite temperature on the behavior of the critical supercurrent is also demonstrated in \ref{sec:appendix A}, in which the height of the plateaus is getting reduced with increasing the value of $T/T_C$ ratio (see Fig. \ref{Fig:9}).

\begin{figure}[]
\begin{center}
\includegraphics[width=4.5in]{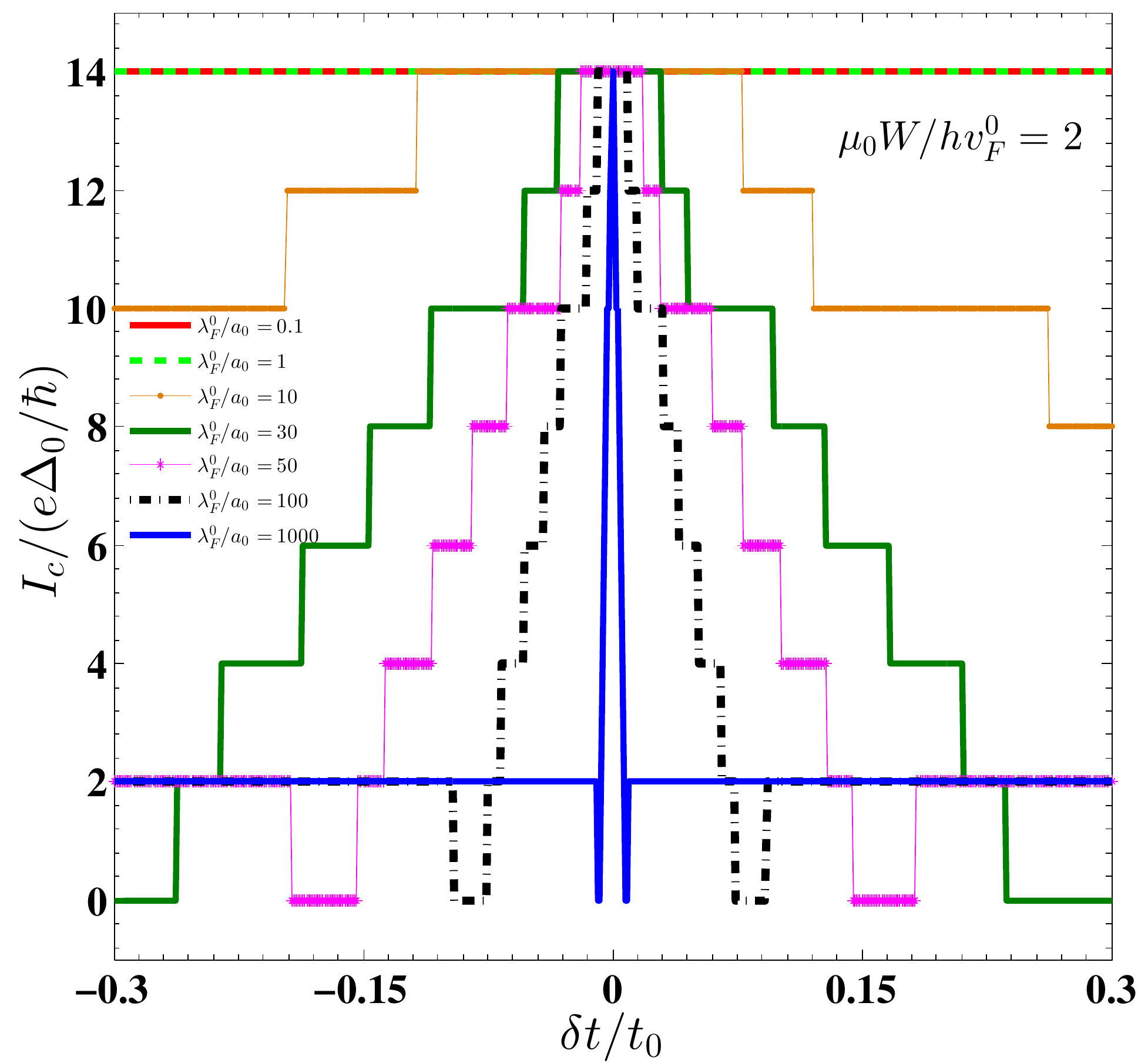}
\end{center}
\caption{\label{Fig:5} The critical supercurrent $I_C$ in terms of the applied strain $\delta t/t_0$ for different values of $\lambda_F^0 / a_0$, when the width of the N region is fixed to $\mu_0 W/h v_{\rm F}^0=2$.  It should be emphasized that a peculiar asymmetric step-like behavior is shown for the critical supercurrent in the case of $\lambda_F^0 / a_0>1$. Moreover, the width of the plateaus and also the asymmetry between the negative and positive $\delta t/t_0$ ratios of $I_C$ are reduced by increasing the $\lambda_F^0 / a_0$ value so that the critical supercurrent reaches a sharp symmetric peak around $\delta t/t_0 =0$ and attains a constant value for finite values of $\delta t/t_0$, when $\lambda_F^0 / a_0$ is large.}
\end{figure}

We further present the behavior of the critical supercurrent in terms of the $\delta t/t_0$ value for different $\lambda_F^0 / a_0$ ratios, when $\mu_0 W/h v_{\rm F}^0 =2$ [see Fig. \ref{Fig:5}]. For small values of $\lambda_F^0 / a_0$, the critical supercurrent has a constant value of $I_C=14 e\Delta_0/\hbar$, independent of the magnitude of applied strain. In the case of $\lambda_F^0 / a_0>1$, a peculiar asymmetric step-like behavior is shown for the critical supercurrent. The supercurrent contains a broad peak with a constant value of around $\delta t/t_0=0$. Increasing the magnitude of the applied strain $|\delta t/t_0|$ leads to decreasing behavior of the critical supercurrent such that it can be switched off for a wide range of $\delta t/t_0$. Note that the plateaus of the critical supercurrent for negative $\delta t/t_0$ ratios are wider than that of the positive $\delta t/t_0$ ratios. This asymmetry in the applied strain $\delta t/t_0$ and also the width of the plateaus are reduced by increasing the $\lambda_F^0 / a_0$ value so that the critical supercurrent reaches a sharp symmetric peak around $\delta t/t_0 =0$ and attains a constant value for finite values of $\delta t/t_0$, when $\lambda_F^0 / a_0$ is large. We have found (not shown) that while the plateaus of the critical supercurrent are getting broader with decreasing the $\mu_0 W/h v_{\rm F}^0$ value, their height is getting reduced. In addition, the critical supercurrent can be switched off for an experimentally available range of $\delta t/t_0$, depending on the value of the Fermi wavelength inside the N region. These results introduce the peculiar quantization of the critical supercurrent in terms of the strain-induced pseudo-magnetic field
[since $\bm{B}_s=(\hbar v_{\rm F}^0)^{-1}\delta t [\delta(x+L/2)-\delta(x-L/2)]\hat{\bm z}$], in contrast to the Fraunhofer pattern of the critical supercurrent in the presence of an applied magnetic field. The absence of the oscillatory supercurrent (Fraunhofer pattern) originates from the fact that strain-induced gauge fields possess opposite directions at the two valleys owing to the time-reversal symmetry. Therefore, there is no net Aharonov-Bohm effect due to the pseudo-magnetic field in the current carried by the bound states composed of electrons and holes from various valleys. The non-Fraunhofer pattern of the supercurrent is also demonstrated in graphene-based Josephson junction under an arc-shaped applied strain~\cite{Khanjani}.

We should mention that the presence of perpendicular pseudo-magnetic field close to the two interfaces inside the strained normal ZGNR, makes the quasiparticles with the excitation energies $\varepsilon_p<\Delta_0$ to have small deviation from their path in the vicinity of the contacts between normal and superconducting regions, immediately before and after undergoing Andreev scattering (at the N/S interfaces). However, the Josephson current will be carried across the junction through Andreev bound states. While in the case of a Josephson junction made of strained graphene~\cite{Peeters} (with finite pseudo-magnetic field which generates pseudo-Landau levels~\cite{Lee}), two regimes are shown for the critical supercurrent depending on the length of the N region. For the junction length smaller than twice the cyclotron radius, quasiparticles will bounce between the superconducting contacts leading to the enhancement of the Josephson current. For long junction regime, quasiparticles will scatter from one superconducting contact thus creating Andreev bound edge states in which case the Josephson current will be strongly suppressed across the junction.

The experimental realization of our theoretical study is quite feasible due to the possibility of creating clean samples of graphene~\cite{Schweizer,Lin}, and successfully inducing superconducting correlations~\cite{Heersche07,Du08,Coskun12,Borzenets11} and loading tensile strain in graphene sheets~\cite{Ni08,Pereira09_1,Lee08,Ferrari,Wang}. As mentioned in Sec. \ref{sec:introduction}, there are several methods which indicate that uniaxial strain has been successfully induced in graphene with the advantage that stretching or geometrical patterning can be made on substrates rather than on graphene. In our case, the uniaxial strain along a prescribed direction can be achieved by depositing the ZGNR on a flexible substrate, like PET (Polyethylene terephthalate) film, and bending the substrate on which graphene is elongated without slippage~\cite{Ni08,Ferrari}. This method provides a way to experimentally control, tune and reproduce the uniaxial strain in graphene by using two- or four-point bending setups~\cite{Ferrari}.

\subsection{Local density of states}\label{sec:LDOS}
\begin{figure}[]
\begin{center}
\includegraphics[width=4.5in]{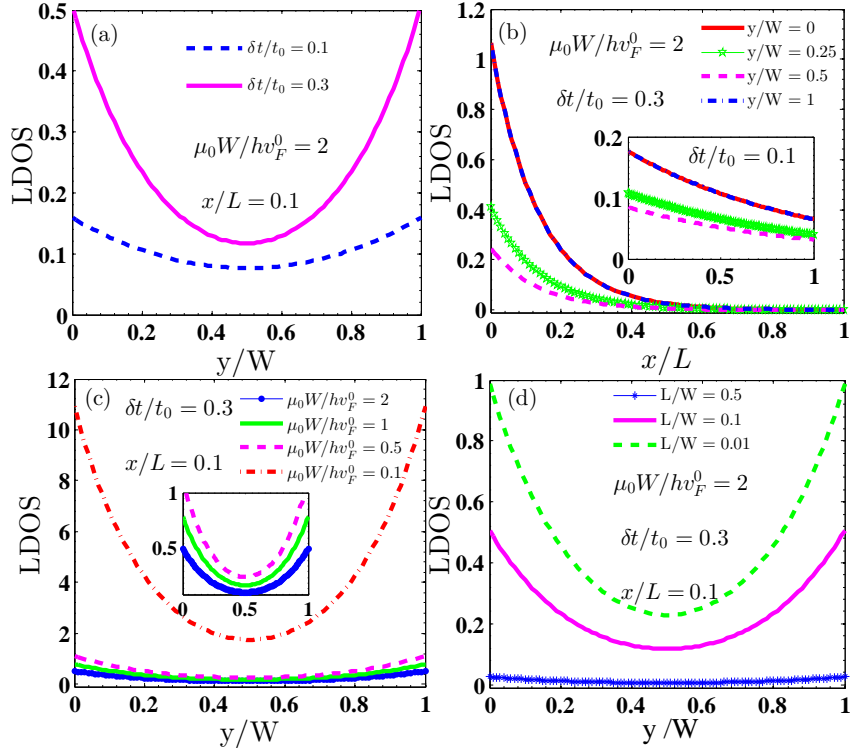}
\end{center}
\caption{\label{Fig:6} Top panel: The local density of states (LDOS) inside the in-plane strained zigzag N nanoribbon versus the dimensionless positions $y/W$ along the y direction for $x/L = 0.1$ (a) and $x/L$ along the x direction for different values of $y/W$ (b), when $\lambda_F^0 / a_0 = 100$, $\mu_0 W/h v_{\rm F}^0=2$, $\varepsilon_p/\Delta_0 = 0$, $L/W = 0.1$ and $\delta t/t_0 = 0.1$ and $0.3$. Bottom panel: The local density of states versus $y/W$ for different values of the width $\mu_0 W/h v_{\rm F}^0$ (c) and the $L/W$ ratio of the N zigzag nanoribbon (d), when $\lambda_F^0 / a_0 = 100$, $x/L=0.1$, $\varepsilon_p/\Delta_0 = 0$, and $\delta t/t_0 = 0.3$. Inset of (b) shows the behavior of the density of states versus $x/L$ for smaller magnitude of the applied strain $\delta t/t_0=0.1$. It should be noted that reducing $\mu_0 W/h v_{\rm F}^0$ and $L/W$ ratios lead to the enhancement of the LDOS especially at the edges of the nanoribbon.}
\end{figure}
\begin{figure}[]
\begin{center}
\includegraphics[width=4.5in]{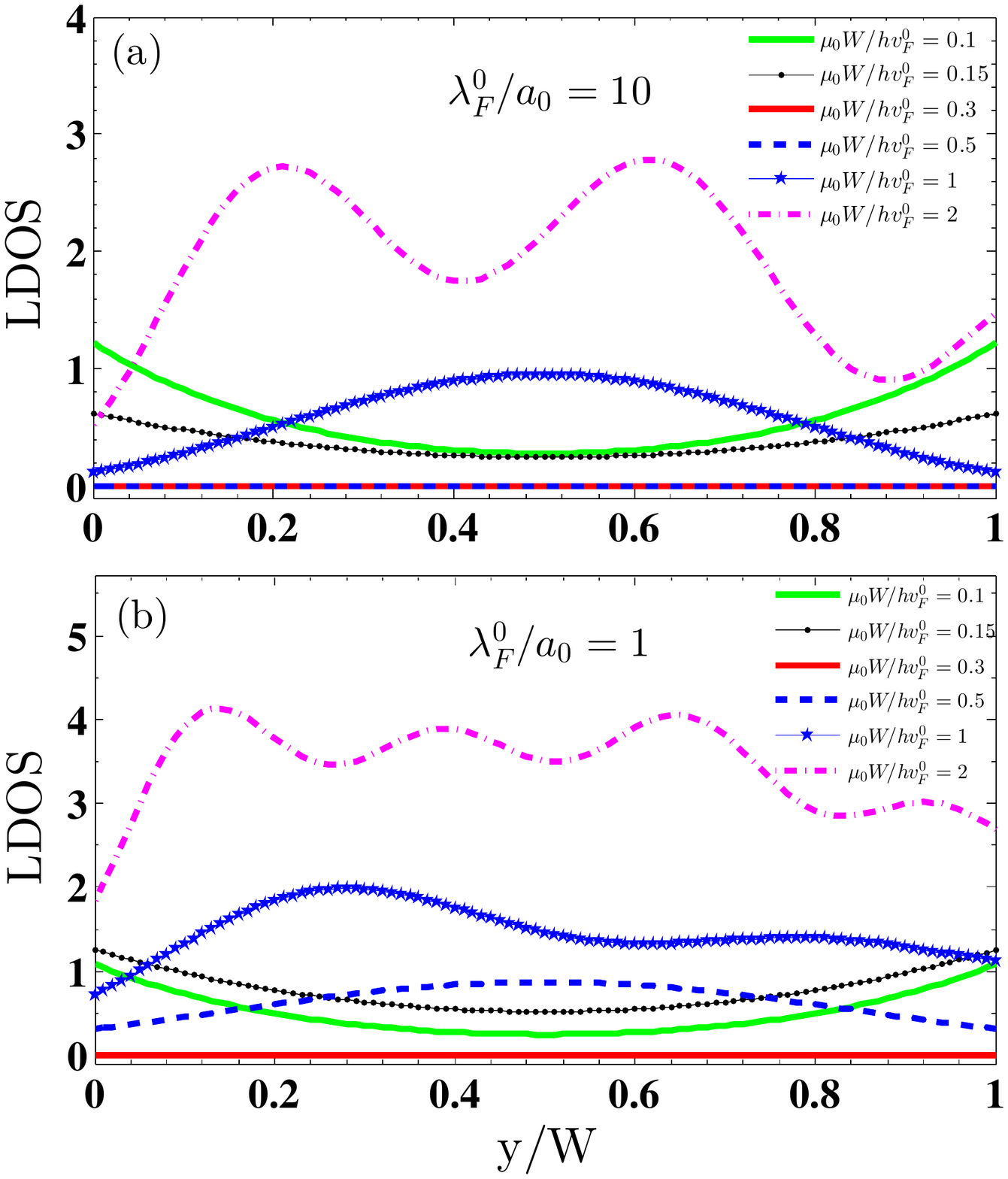}
\end{center}
\caption{\label{Fig:7} Top (bottom) panel: The behavior of the local density of states versus $y/W$ for different values of $\mu_0 W/h v_{\rm F}^0$, when $\lambda_F^0 / a_0=10$ ($\lambda_F^0 / a_0 = 1$), $\varepsilon_p/\Delta_0 = 0$, $\delta t/t_0 = 0.3$, $x/L=0.1$ and $L/W=0.1$. The position dependence of the LDOS is very sensitive to the value of $\lambda_F^0 / a_0$ as well as $\mu_0 W/h v_{\rm F}^0$. In contrast to Fig. \ref{Fig:6}(c), increasing the width of the N region makes the LDOS to change its behavior from damping to oscillating along the $y-$direction and be constant along the $x$-direction when the Fermi wavelength of the N region is reduced to $\lambda_F^0 / a_0=10$ (a) and $\lambda_F^0 / a_0=1$ (b). The crossover from damping to oscillatory behavior occurs at lower values of the $\mu_0 W/h v_{\rm F}^0$ ratio when the $\lambda_F^0 / a_0$ ratio is reduced to unity.}
\end{figure}
\begin{figure}[]
\begin{center}
\includegraphics[width=4.5in]{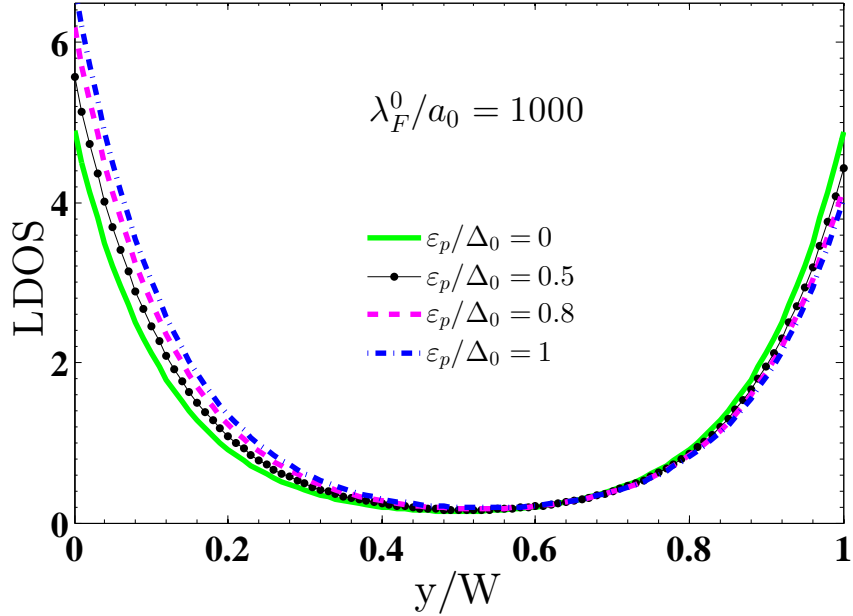}
\end{center}
\caption{\label{Fig:8} The behavior of the LDOS versus the dimensionless position $y/W$ for different values of the excitation energy, when $\lambda_F^0 / a_0 = 1000$. We set $\mu_0 W/h v_{\rm F}^0=2$, $\delta t/t_0 =0.3$, $x/L=0.1$ and $L/W=0.01$. Note that increasing the $\varepsilon_p/\Delta_0$ ratio results in an enhancement of the LDOS at the edge $y=0$ and reduction of its value at the opposite edge $y=W$.}
\end{figure}

Now, we proceed to investigate the behavior of the LDOS inside the in-plane strained N graphene nanoribbon with zigzag edges. We denote the local density of states $\tilde{N}(E,r)$ by LDOS in Figs. \ref{Fig:3}(c), \ref{Fig:6}, \ref{Fig:7} and \ref{Fig:8}. Figures \ref{Fig:6}(a) and \ref{Fig:6}(b) respectively present the behavior of the LDOS in terms of the dimensionless distances $y/w$ and $x/L$ along the $y$ and $x$ directions, when $\lambda_F^0 / a_0 = 100$, $\varepsilon_p/\Delta_0 = 0$, $\mu_0 W/h v_{\rm F}^0=2$, $L/W = 0.1$ and $\delta t/t_0 = 0.1$ and $0.3$. The high low-energy density of states is shown at opposite edges of the zGNR resulting from the $y-$dimensional evanescent edge mode (imaginary $q$) which is being decreased by going away from the edges [see Fig. \ref{Fig:6}(a)]. Moreover, the LDOS undergoes a decaying behavior with a distance of $x/L$ from the S/N interface [Fig. \ref{Fig:6}(b)]. It is indicated that increasing the magnitude of the applied strain leads to the enhancement of the LDOS inside the N region. Therefore in the following figures, we set $\delta t/t_0 = 0.3$. Moreover, the effects of the width $\mu_0 W/h v_{\rm F}^0$ and $L/W$ ratios of the strained nanoribbon on the LDOS are investigated in Figs. \ref{Fig:6}(c) and \ref{Fig:6}(d). It is worth mentioning that reducing $\mu_0 W/h v_{\rm F}^0$ and $L/W$ ratios lead to the enhancement of the LDOS especially at the edges of the nanoribbon.

Most importantly, we find the position dependence of the LDOS is very sensitive to the value of the Fermi wavelength $\lambda_F^0 / a_0$ ratio as well as the width of the strained N nanoribbon $\mu_0 W/h v_{\rm F}^0$ [see Fig. \ref{Fig:7}]. In contrast to large $\lambda_F^0 / a_0$ ratio [Fig. \ref{Fig:6}(c)], increasing the width of the N region makes the LDOS to change its behavior from damping to oscillating along the $y-$direction and be constant along the $x$-direction when the Fermi wavelength of the N region is reduced to $\lambda_F^0 / a_0=10$ [Fig. \ref{Fig:7}(a)] and $\lambda_F^0 / a_0=1$ [Fig. \ref{Fig:7}(b)]. The crossover from damping to oscillatory behavior occurs at lower values of the $\mu_0 W/h v_{\rm F}^0$ ratio when the $\lambda_F^0 / a_0$ ratio is reduced to unity. In addition, the amplitude and period of the oscillations (resulted from the propagating modes) decrease with reducing the $\lambda_F^0 / a_0$ ratio and the local density of states at the edge $y=w$ has larger value compared to that of the opposite edge $y=0$. Note that according to the boundary condition of zigzag edges applied to the strained N nanoribbon (see Sec. \ref{sec:model}), the states on the edges with the position $y=0$ and $y=W$ are those respectively coming from the A and B sublattices.

To explain the behavior of the LDOS, we look at the squared absolute value of the wave function of the state with zero excitation energy, $|\psi_{\bm{k}}(r)|^2$ (see \ref{sec:appendix B}). It is shown that the oscillating LDOS results from the propagating state with transcendental relation $\sin[{qW\pm(\alpha+\alpha')/2}]=0$ for transverse momenta in the range $|q|\leq(\mu-\delta t)/\hbar v_{\rm F}$ [see Eq. (\ref{psiN,prop})]. While the decaying LDOS originates from the evanescent edge state with transcendental relation $\sinh[{(b+b')/2}]=\pm\sinh{(qW)}$ for transverse momenta $q$ [see Eq. (\ref{psiN,evan})]. In the case of $\delta t\ll \mu$ (with
 $\delta t/\mu=(\delta t/t_0)(\lambda_F^0 / a_0)\sqrt{1+\epsilon_{xx}+\epsilon_{yy}}/3\pi[1+(1-\beta)(\epsilon_{xx}+\epsilon_{xy})]$), the transcendental relations of quantized $q$ become $\sin{(qW)}/qW=\pm \hbar v_{\rm F}/\mu W$ and $\sinh{(qW)}/qW=\pm \hbar v_{\rm F}/\mu W$ for the propagating and evanescent modes, respectively. This implies the oscillating regime to exist for the range of $\mu W/\hbar v_{\rm F}\geq 1$ and the decaying regime for the range of $\mu W/\hbar v_{\rm F}<1$.

We further demonstrate the dimensionless position $y/W$ dependence of the LDOS for different values of the excitation energy in Fig. \ref{Fig:8}, when $\lambda_F^0 / a_0=1000$. Similar to the case of $\lambda_F^0 / a_0=100$ [see Fig. \ref{Fig:6}(a)], the LDOS has high equal values at the two edges and decreases monotonically with the distance from the edges, when $\varepsilon_p/\Delta_0=0$. We find that finite excitation energy makes the LDOS to have different values at the two edges of the zGNR. Increasing the value of the $\varepsilon_p/\Delta_0$ ratio leads to the enhancement of the LDOS at the edge $y=0$ and reduction of its value at the opposite edge $y=W$.

For the sake of completeness, we present the behavior of the local density of states in terms of the $\mu_0 W/h v_{\rm F}^0$ ratio for three different values of the dimensionless distance from the edges $y/W = 0, 0.5$ and $1$ in Fig. \ref{Fig:3}(c), when $x/L = 0.1$, $L/W = 0.1$, $\varepsilon_p/\Delta_0=0$, $\delta t/t_0 = 0.3$ and $\lambda_F^0/a_0 = 5$. It is figured out that the appearance of a flat band located at small $\mu_0 W/h v_{\rm F}^0$ value, gives rise to a strong LDOS peak at the edges of the strained zGNR with $y/W =0$ and $1$. Especially, the quantized level singularities, finger print of the supercurrent plateaus, are clearly observed in the density of states accompanied with the vanishing LDOS for the values of the $\mu_0 W/h v_{\rm F}^0$ ratio being inside the gray region of Fig. \ref{Fig:3}(a).

\section{Conclusion}\label{sec:conclusion}
In conclusion, we have studied the Josephson effect in a short strained graphene nanoribbon with zigzag edges and an arbitrary width of $W$ connecting two superconducting electrodes. We have found that this type of edges supports an integer quantization of the supercurrent to even multiples of the supercurrent quanta ($e\Delta_0/\hbar$) in an ordinary superconducting quantum point contact. The strain-induced plateaus placing between the steps of the unstrained structure have a small width, which can be enhanced by increasing the strength of the applied strain and the Fermi wavelength inside the pristine graphene nanoribbon, and also they can be shifted to higher values of $\mu_0 W/h v_{\rm F}^0$ ($\mu_0$ and $v_{\rm F}^0$ are respectively the chemical potential and Fermi velocity inside the pristine graphene nanoribbon). Also, the height of the plateaus can be reduced by increasing the temperature below the critical temperature of the superconducting order parameter. We have further demonstrated a particular asymmetric step-wise variation of the critical supercurrent versus the strain-induced pseudo-magnetic field inside the graphene nanoribbon in which the height and width of the steps can be decreased by increasing the strength of the fictitious gauge potential (or strain) and the Fermi wavelength, respectively. In addition, the results show that the Josephson current can be turned on/off in presence of the applied strain, depending on the Fermi wavelength and width of the normal nanoribbon. We claim that our proposed superconducting quantum point contact can be utilized as a supercurrent switch.

We have further studied the behavior of the local density of states inside the strained normal nanoribbon with zigzag edges. We have clearly shown the high density of states with equal magnitudes at the opposite edges of the nanoribbon resulting from the evanescent edge mode when the excitation energy is zero. The finite excitation energy leads to the enhancement of the local density of states at the edge $y=0$ and reduction of its value at the other edge $y=W$. The local density of states has a decaying behavior with the distance from the edges for large values of the Fermi wavelength inside the pristine normal region and importantly it undergoes an oscillatory behavior with different values at the two edges for smaller values of the Fermi wavelength, by tuning the width $\mu_0 W/h v_{\rm F}^0$ ratio. Moreover, we have demonstrated that the local density of states increases by enhancing the magnitude of the applied strain as well as reducing the length to width ($L/W$) ratio of the graphene nanoribbon.

\section*{Acknowledgments}

L. M is supported by Iran Science Elites Federation and R. A was supported by the Australian Research Council Centre of Excellence in Future Low-Energy Electronics Technologies (project number CE170100039).

\appendix
\section{\label{sec:appendix A}Finite temperature behavior of the critical supercurrent}

The results of the finite-temperature behavior of the critical supercurrent $I_C$ with respect to the width of the normal region $\mu_0 W/h v_{\rm F}^0$ are presented in Fig. \ref{Fig:9}. Making use of relation $\Delta_0=1.76\ k_B T_C$, the temperature $T$ is parameterized in units of the critical temperature of the superconducting order parameter $T_C$. We set the Fermi wavelength $\lambda_F^0 / a_0 = 5$ and the applied strain ratio $\delta t/t_0 = 0.3$. The peculiar step-wise variation of the critical supercurrent is preserved for finite-temperatures below the critical temperature $T_C$. Comparing the results with that of the zero temperature (Fig. \ref{Fig:2}) demonstrates the reduction of the critical supercurrent as well as the height of the steps by growing the temperature $T/T_C$.

\section{\label{sec:appendix B}Squared absolute value of the wave function inside the strained zigzag-edge normal nanoribbon}
To evaluate the behavior of the local density of states inside the strained normal nanoribbon with zigzag edges by making use of Eq. (\ref{LDOS}), we need to calculate the contribution of both propagating and evanescent modes to the squared absolute value of the wave function,
\begin{equation}
|\psi_{\bm{k}}(r)|^2=|\psi_{\bm{k}}^{prop}(r)|^2+|\psi_{\bm{k}}^{evan}(r)|^2.
\end{equation}
The contribution of the propagating modes is written as
\begin{equation}
|\psi_{\bm{k}}^{prop}(r)|^2=|\psi_{N}^{prop}|^2+|\psi_{N}^{'prop}|^2,
\end{equation}
with
\begin{equation}
\label{psiN,prop}
|\psi_{N}^{(')prop}|^2=2\sin^2({qy})+2\sin^2[{qy\pm(\frac{\alpha+\alpha'}{2})}],
\end{equation}
for incoming electrons from the K(K') valley with the wave function $\psi_{N}^{(')prop}$ and the transcendental relation $\sin[{qW\pm(\alpha+\alpha')/2}]=0$ for transverse momenta in the range $|q|\leq(\mu-\delta t)/\hbar v_{\rm F}$. The contribution of the non-zero evanescent modes (edge modes) read as
\begin{equation}
|\psi_{\bm{k}}^{evan}(r)|^2=|\psi_{N}^{evan}|^2+|\psi_{N}^{'evan}|^2,
\end{equation}
with
%\begin{widetext}
\begin{eqnarray}
\label{psiN,evan}
|\psi_{N}^{(')evan}|^2&=&\sinh^2({qy})\ [e^{\mp2Im(k)x}+e^{\pm2Im(k')x}]+\frac{1}{2} [\cosh{(2qy\pm b\pm b')}\nonumber\\
&-&\cos{(a+a')}]\ [e^{\mp(2Im(k)x+b-b')}+e^{\pm(2Im(k')x+b-b')}],
\end{eqnarray}
%\end{widetext}

where $\psi_{N}^{evan}$ and $\psi_{N}^{'evan}$ are respectively the wave functions of the incoming electrons from the K(K') valley with the transcendental relation $\sinh{[(b+b')/2]}=\mp\sinh{(qW)}$ for transverse momenta $q$.

The parameters in Eqs. (\ref{psiN,prop}), and (\ref{psiN,evan}) are defined by $k^{(')}=\mu\ [\cos{a}^{(')}\cosh{b}^{(')}-i\ \sin{a}^{(')}\sinh{b}^{(')}]/\hbar v_{\rm F}$, $a^{(')}=-atan2\ [(\delta ^{(')}\mp \delta t/\mu),(\gamma^{(')}-\hbar v_{\rm F} q/\mu)]$, $b^{(')}=Log\ [\sqrt{(\gamma^{(')}-\hbar v_{\rm F} q/\mu)^2+(\delta ^{(')}\mp \delta t/\mu)^2}]$, $\gamma^{(')}=\sqrt{[\eta+\sqrt{\eta^2+{\chi^{(')}}^2}]/2}$, $\delta^{(')}=sgn(\chi^{(')})\sqrt{[-\eta+\sqrt{\eta^2+{\chi^{(')}}^2}]/2}$,  $\eta=1-(\delta t/\mu)^2+(\hbar v_{\rm F} q/\mu)^2$, and $\chi^{(')}=\pm 2(\delta t/\mu)(\hbar v_{\rm F} q/\mu)$. Note that the above-found relations are for the case of zero excitation energy.

\end{document}